\documentstyle[psfig,conf_gemini,10pt]{article}
\begin{document}
\def\Sec{${}^{\prime\prime}$}
\heading{%
%
\centerline{\bf A NIR search for high-redshift clusters}\\
%
} 
\par\medskip\noindent
\author{%
C. Mendes de Oliveira$^{1,2}$, U. Hopp$^{2}$, R. Bender
$^{2}$, N. Drory$^{2}$, R.P. Saglia$^{2}$
}

\address{%
Instituto Astron\^omico e Geof\'{\i}sico, Av Miguel St\'efano 4200,
04301-904, S\~ao Paulo, Brazil
}
\address{%
Universit\"ats-Sternwarte, Ludwig-Maximilians-Universit\"at,
Scheinersstrasse 1, 81679 Munich, Germany
}
%

{\bf Abstract:} We describe our preliminary results on a search for
high-redshift ($z >$ 0.5) galaxy clusters using near-infrared
photometry obtained with the Omega camera at the 3.5m telescope at
Calar Alto.

\bigskip
\medskip
\centerline{\bf 1. Introduction}
\medskip\noindent 

  The project described here is a search for high-redshift clusters
using VRIJK photometry. We have four main goals:  1) determine the
space density of $z >$ 1 clusters; 2) measure large-scale structure at
redshifts $z >$ 0.5; 3) test the number density evolution of elliptical
galaxies; 4) select a sample of large--$z$ elliptical galaxies for
populations studies.  (1), (2) and (3) provide strong discrimination of
different cosmologies, (4) will give a time baseline to test models of
elliptical galaxy formation and evolution.

{\bf Why do we expect to find clusters at $z >$ 1?} There are several
lines of evidence that suggest there has been little evolution between
$z\sim 1$ and the present.  The passive evolution of
the fundamental plane relations of
cluster galaxies out to $z=0.6$, the small-scatter of the
colour-magnitude diagram and Mg--$\sigma$ relation 
for medium-redshift clusters and the mild
change in the mean colour of cluster elliptical galaxies with redshift
out to $z=1$ (Kelson et al. 1997, Ellis et al. 1997, Bender et al.
1996, 1998, Ziegler \& Bender 1997, Stanford et al.
1998) are some of the evidences which point to a redshift of formation
of cluster ellipticals larger than 2.

  A few examples of high-redshift clusters have already been found,
e.g. 3C 324 at $z=$1.20 (Dickinson 1995), CIG J0848+4453 at z=1.27
(Stanford et al. 1997), but no systematic search using NIR colours
and covering a large area on the sky has yet been done.

  Near-infrared imaging surveys offer one of the most useful tools for
finding clusters at high redshifts.  Other successful techniques are
based on the detection of optical background fluctuations over hundreds
of square degrees on the sky  (Zaritsky et al.  1997) and searches using
ROSAT x-ray observations (e.g. Rosati et al. 1995 -- these are mainly 
restricted to $z<$ 1 clusters however).

{\bf Why search for clusters using NIR photometry?} Old,
passively-evolving galaxies observed at high-z have very red
optical-NIR colours. It is then natural to look for them with very red
filters.  Identifying high-redshift galaxies using NIR photometry
avoids confusion with galactic stars or blue field galaxies.  In
addition, NIR colours are weakly affected by evolution. There is also
the advantage of a small and uniform k-correction for all galaxy
types.

\medskip\bigskip
\centerline{\bf 2. Previous NIR galaxy surveys}
\medskip\noindent

  The largest faint galaxy photometric survey in the near-infrared is 
the ``ESO K'-band survey''.  $B$ and $K^\prime$ images were obtained over 40
arcmin$^2$ down to $K^\prime=20$ and 170 arcmin$^2$ to $K^\prime =19$.
A recent paper on the galaxy counts was published by Saracco et al.
(1997). They identified a population of blue objects
($B-K^\prime <3$) brighter than $K^\prime =18$ which they suggest represent a
population of sub-L$^*$ nearby evolving galaxies.

  Another large galaxy survey which includes photometry in the
near-infrared is that of Elston, Eisenhardt \& Stanford (1995).  They
performed a BRIzJK field survey over 100 arcmin$^2$ down to K=22 and
found one cluster at z=1.27 (later confirmed by Keck spectroscopy of 8
galaxies, Stanford et al. 1997).

  There are several other deep NIR surveys which, however, cover very small
areas from 1 to 20 arcmin$^2$ (eg. Cowie et al. 1990, Cowie et al.
1994, Soifer et al.  1994, McLeod et al. 1995; Djorgovski et al. 1995,
Moustakas et al.  1997 and others). The main goal of these surveys was
to study galaxy evolution through number counts in the K band.

  Optical/near-infrared surveys designed to look for rich clusters of
galaxies at high redshifts have been described by Dickinson and
collaborators.  Imaging of the environment of powerful radio galaxies
yielded the identification of several candidate rich
clusters, visible as an overdensity of red galaxies, some of which were
confirmed to be clusters with subsequent spectroscopy (Dickinson 1995).

\medskip\bigskip \centerline{\bf 3. Our survey} \medskip\noindent

J and $K^\prime$ images were taken at the 3.5m-telescope at Calar Alto
targeting $z >$ 0.5 quasars.  The NIR camera (Omega) with a Rockwell
1024 x 1024 pix (HgCdTe detector) giving  0.4"/pixel and a field of
view of 6.8' x 6.8' was used.  Typical exposure times were 20 min. in J
and $K^\prime$.  About 20 fields were observed ($\sim 46$ arcmin$^2$ each)
covering 0.26 degree$^2$ in total.  Our goal is to cover a 3 degree$^2$
field.

 Seven of the fields have been processed so far. These sample the
environment of quasars with redshifts 0.5 $<$ $z$ $<$ 2.0. We give some
of our preliminary results based on the J and $K^\prime$ images of
these fields in the next section.  A publication in preparation will
describe in more detail the results for the whole sample.  Optical
photometry on these fields is planned for the near future.

\begin{figure}
\centerline{\vbox{
\psfig{figure=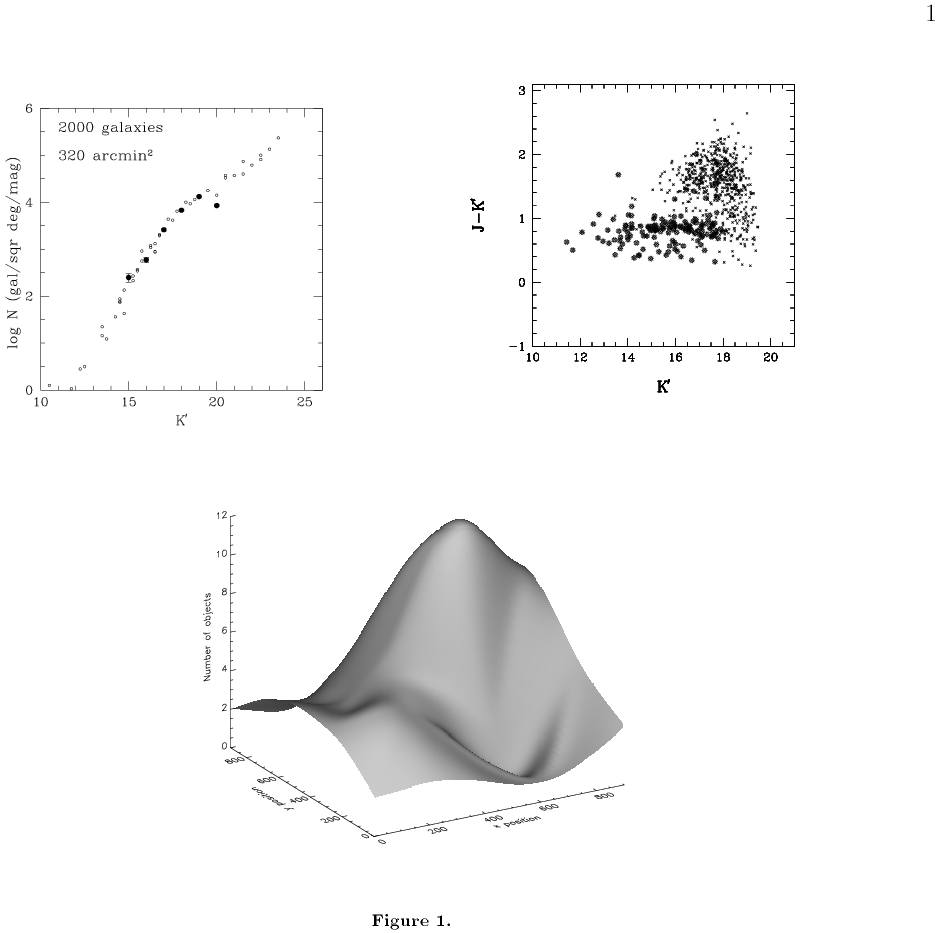,height=17cm}
}}
\caption[]
{{{\bf 1a)}  K$^\prime$ counts corrected for incompleteness
for the seven fields covering an area of 320
sq. arcmin (black dots). 
Overplotted are data from the literature (open symbols). 
{\bf 1b)} K$^\prime$ vs.
J-K$^\prime$ colour for five of the fields (starred symbols are objects with
stellar profiles, small crosses are
galaxies)
{\bf 1c)}
Counts of red galaxies (J-K$^\prime >$ 1.7)
in circles of one arcmin radius centered on
16 positions on one of our fields.
There is a 3$\sigma$ surface overdensity of red galaxies
around the location of the quasar at $z$ = 0.72 (x=600, y=600). }}
\end{figure}

\medskip\bigskip
\centerline{\bf 4. Preliminary results}
\medskip\noindent

    Detection and photometry of the objects in the images was done
using the software package SEXTRACTOR (Bertin \& Arnouts 1996).
Galaxy/star separation was possible to magnitude $K^\prime =18$.
Add-star experiments were performed to assess the photometric
completion.  Our results show that the raw $K^\prime$ counts are 68\%
complete at $K^\prime$=19.5.  In Fig. 1a we show the $K^\prime$ counts
for the seven fully reduced fields, compared to the
$K^\prime$ counts obtained from data published previously in the
literature (Djorgovski et al. 1995, Glazebrook et al. 1994,
Gardner et al. 1993). About 2000 galaxies were detected in an area of
320 arcmin$^2$ in our survey. The area we have covered so far is larger
than that for the largest previously published survey (to the magnitude
limit we reach).  Fig. 1a shows the number of galaxies per degree$^2$,
per magnitude bin.  The shape of the curve is very similar to that
found in previous studies. In particular, we confirm the change of
slope of the curve at a magnitude of $K^\prime \sim 17$.

   $J-K^\prime$ colours were determined for all objects detected in $K^\prime$.
Colours within an aperture of diameter 4 arcsec vs. $K^\prime$ magnitudes are
plotted in Fig. 1b for five of the fields for which $J$ and $K^\prime$ images
were available.  It is clear that there are a number of galaxies which
are as red as early-type galaxies at $z \sim 1.2$ ($J-K^\prime > 1.8$). We find
about 1 galaxy with $J-K^\prime >1.8$ per arcmin$^2$ to a limiting magnitude
$K^\prime$=19 mag.

    We performed a search for a surface density enhancement of red
objects, as a first indication of the presence of candidate clusters in
our fields.  In only one field was such an enhancement found.  Fig. 1c
shows a plot of the number of galaxies with magnitudes in the interval
$K^\prime =$ 17--18.5, with $J-K^\prime >$ 1.7 in circles of 1 arcmin
radius centered at 16 positions for one of the frames. We find a
3$\sigma$ surface overdensity of red galaxies around a quasar at
redshift $z$ = 0.72 (position 600, 600) which we explain as a possible
poor cluster of galaxies.  Optical VRI images of the candidate cluster
and preliminary photometric redshift determinations (described in a
paper in preparation) confirmed the presence of a density enhancement
in this area. Subsequent spectroscopy is planned to confirm the nature
and redshift of the cluster.
\medskip\bigskip

\centerline{\bf Acknowledgements}
\medskip\noindent
We thank the staff at Calar Alto for helping with the observations.
CMdO acknowledges the financial support from the Alexander von Humboldt
Foundation. This work was supported by the Sonderforschungsbereich
SFB375.


%
\vfill
\end{document}